\documentclass[twocolumn]{aastex631}

\usepackage{amsmath}

\shorttitle{Particle Acceleration in Relativistic Perpendicular Shocks with Proton Admixture}
\shortauthors{Yu et al.}
\graphicspath{{./}{figures/}}

\begin{document}

\title{Kinetic Simulations of Particle Acceleration in Relativistic Perpendicular Electron-positron Shocks with Proton Admixture}

\correspondingauthor{Jun Fang}
\email{fangjun@ynu.edu.cn}

\author{Huan Yu}
\affiliation{Department of Physical Science and Technology, Kunming University, Kunming 650214, China}
\author{Qi Xia}
\author[0000-0001-8043-0745]{Jun Fang}
\affiliation{Department of Astronomy, School of Physics and Astronomy, Key Laboratory of Astroparticle Physics of Yunnan Province, Yunnan University, Kunming 650091, China}

\begin{abstract}
Particle acceleration in relativistic shocks of electron-positron plasmas with proton admixture is investigated through two-dimensional (2D) particle-in-cell (PIC) simulations. The upstream plasma, with a bulk Lorentz factor of $10$ and a magnetization parameter of 0.02, includes a small fraction of protons ($\sim 5\%$ by number). A relativistic perpendicular shock is formed by reflecting the flow off a conducting wall. The shock structure, electromagnetic fields, and particle energy spectra are analyzed. The particle density and the magnetic field have fluctuations. In the far-downstream region of the shock, positrons are accelerated to energies comparable to protons and develop a hard nonthermal component with a spectral index of $\sim 2$ in their energy spectrum, while electrons remain confined to lower energies. This asymmetry is attributed to the polarization properties of proton-driven electromagnetic waves, which favor positron acceleration. The results highlight the importance of plasma composition in shaping particle acceleration and nonthermal emission in relativistic shocks. These findings provide new insights into the microphysics of particle acceleration in astrophysical sources containing relativistic shocks.
\end{abstract}


\section{Introduction}
\label{sec:Intro}

Relativistic shocks are among the most powerful particle accelerators in the universe, playing a central role in energizing nonthermal emission across a wide range of astrophysical objects such as pulsar wind nebulae (PWNe), active galactic nuclei (AGN) jets, gamma-ray bursts (GRBs) \citep{KC84, MG85, P04, Mea16, BT11}. In these systems, relativistic shocks convert the kinetic energy of the outflow into thermal and nonthermal particle energy, and the detected multiband emission is produced by the accelerated particles \citep{FZ10,Tea14,Zea18,Xea19}.

Particle acceleration in relativistic collisionless shocks is a fundamental process in high-energy astrophysics, powering nonthermal emission in various astrophysical sources. Recent kinetic simulations have provided significant insights into the mechanisms of particle acceleration for the shocks in different plasma compositions and magnetic field configurations \citep{S08,Mea09,SS09,Lea19,Fea19,Xea20,KR21,Fea22,Gea24,Guea24}.

\citet{S08} demonstrated that relativistic shocks propagating in unmagnetized electron-positron plasmas can self-consistently accelerate particles through a Fermi-like process. Their long-term 2D PIC simulations revealed the development of a high-energy power-law tail in the downstream particle spectrum, with an index of $\sim -2.4$. The acceleration mechanism was attributed to particles bouncing between upstream and downstream regions in self-generated magnetic fields produced by the Weibel instability. This work established that at least $\sim1\%$ of downstream particles are accelerated into the nonthermal tail, carrying $\sim10\%$ of the downstream kinetic energy. In their study on the long-term evolution of relativistic unmagnetized collisionless shocks, \citet{Gea24} conducted 2D PIC simulations of unprecedented duration and size to investigate the dynamics of electron-positron shocks propagating in an unmagnetized ambient medium. They observed the generation of intermittent magnetic structures that grow in size over time, reaching scales of approximately $100$ plasma skin depths by the end of the simulation.

For magnetized plasmas, \citet{SS09} investigated particle acceleration in relativistic magnetized pair shocks using PIC simulations. They found that efficient particle acceleration occurs only in subluminal shocks, where the magnetic field inclination angle with respect to the shock normal is less than a critical angle. The downstream spectrum in such shocks consists of a relativistic Maxwellian and a high-energy power-law tail, with the acceleration efficiency increasing with magnetic obliquity. For nearly parallel shocks, diffusive shock acceleration dominates, while shock-drift acceleration becomes more important at larger subluminal inclinations. Their results place important constraints on models of PWNe, GRBs, and AGN jets that invoke particle acceleration in relativistic magnetized shocks.

The presence of heavier ions, such as protons, can significantly alter the particle acceleration in shocks. Protons, due to their larger mass, can drive cyclotron instabilities that generate high-harmonic ion cyclotron waves. These waves can be absorbed by pairs, providing a mechanism for nonthermal acceleration \citep{Hea92, AA06, Yea24}. One-dimensional (1D) PIC simulations have revealed key aspects of nonthermal acceleration in electron-positron-proton plasmas \citep{AA06}. When protons constitute a non-negligible fraction of the plasma, the efficiency of nonthermal acceleration increases. Positrons often receive a larger share of the nonthermal energy than electrons since the left-handed nature of the waves generated by the ions enables efficient positron energization. However, higher-dimensional simulations are essential for capturing the full complexity of shock structures and particle acceleration processes. \citet{Sea12} performed 2D PIC simulations of relativistic perpendicular shocks in electron-positron-proton plasmas, focusing on the role of proton-driven instabilities in particle acceleration. They found that the presence of protons significantly enhances the efficiency of nonthermal acceleration, particularly for positrons, which are accelerated to energies comparable to those of protons.

In this work, we present 2D PIC simulations of relativistic shocks in electron-positron plasmas with a small admixture of protons ($\sim 5\%$ by number). Our results extend previous findings and provide new insights into the mechanisms underlying the nonthermal emission observed in astrophysical sources. In Section \ref{sec_simu}, the simulation setup for the relativistic shock is described. The shock structure and the particle spectra are present in Section \ref{sec_result}. The conclusions and some discussion are given in Section \ref{sec_cd}.

\section{Simulation Setup}
\label{sec_simu}
The simulation setup is designed to study the formation and evolution of relativistic perpendicular shocks in a plasma composed of electrons, positrons, and a small fraction of protons. The choice of a perpendicular shock geometry is motivated by its relevance to astrophysical environments such as PWNe, whose termination shocks are often derived to be perpendicular from MHD simulations \citep{Pea14,Oea16}. The PIC simulations on the perpendicular shock are performed using the \texttt{Smilei} code \citep{Dea18}, which is widely used for studying plasma physics in relativistic regimes. Below, we provide a detailed description of the simulation parameters, initial conditions, and the physical rationale behind the choices made in the setup.

The simulation domain is initialized with a plasma moving in the $-x$ direction with a bulk Lorentz factor of $\gamma_0 = 10$. The charge-neutral plasma consists of electrons ($e^-$), positrons ($e^+$), and protons ($p$), and the number densities for them are $n_{e^-} = 1.1n_0$, $n_{e^+} = n_0$, and $n_p = 0.1n_0$, respectively. The small fraction of protons is chosen to reflect the typical composition of astrophysical plasmas where protons, though less abundant than electrons and positrons, play a significant role in the dynamics of the shock. The reduced mass ratio between protons and electrons is set to $m_p / m_e = 30$, which is a compromise between computational feasibility and physical realism.  The kinetic energy contained in protons accounts for $58.8\%$ of the total kinetic energy of the plasma. While the actual mass ratio in nature is $m_p / m_e \approx 1836$, the reduced ratio is commonly used in PIC simulations to reduce computational costs while still capturing the essential physics. The temperature of all species is set to a non-relativistic value ($k_{\mathrm{B}} T =2\times10^{-3} m_e c^2$) to approximate a cold plasma. This choice ensures that the initial thermal energy of the particles is negligible compared to their kinetic energy, allowing us to focus on the effects of the shock acceleration process.

A uniform magnetic field $\mathbf{B}_0$ and an electric field $\mathbf{E_0=-\beta_0\hat{x}\times\mathbf{B_0}}$ ($\beta_0 = (1-1/\gamma_0^2)^{1/2}$) are imposed along the $z$ axis and the $y$ axis, respectively, defining the perpendicular shock geometry. The magnetization parameter $\sigma$, defined as the ratio of magnetic energy density to particle kinetic energy density, is set to $\sigma = B_0^2 / (8 \pi \gamma_0(n_{e^-}m_e + n_{e^+}m_e + n_pm_p)c^2)=2.0 \times 10^{-2}$ in the simulations, and the Alfv\'{e}n Mach number is $M_{\mathrm{A}} \approx ((1+\sigma)/\sigma)^{1/2} = 7.1$ for the relativistic shock.

The simulation domain spans a rectangular region in the $x$-$y$ plane with an extension of $12288 \, c/\omega_{\mathrm{pe}} \times 128 \, c/\omega_{\mathrm{pe}}$, where $\omega_{\mathrm{pe}} = \sqrt{4 \pi n_e e^2 /\gamma_0 m_e}$ is the  relativistic electron plasma frequency, with periodic boundary conditions applied in the $y$ direction. The relativistic Larmor frequency of the upstream protons is $\Omega_{\mathrm{Lp}} = e B_0/\gamma_0 m_{\mathrm{p}} c = 6.67\times10^{-3} \omega_{\mathrm{pe}}$, and the relativistic proton Larmor radius is $r_{\mathrm{Lp}} \equiv \gamma_0 \beta_0 m_{\mathrm{p}} c^2 /(eB_0) = 4.72\times10^2\,c/\omega_{\mathrm{pe}}$.
The resolution of the grid is $\Delta x = \Delta y = 1/8\, c/\omega_{\mathrm{pe}}$, ensuring that the electron skin depth is well-resolved. Each cell is populated with  $16$ particles per species and the time step is set to $\Delta t = 1/32\ \omega_{\mathrm{pe}}^{-1}$. Temporal Friedman smoothing ($\theta=0.1$) and binomial current filtering (three passes) are used to suppress artificial numerical artifacts, such as spurious upstream heating, caused by grid-Cherenkov instability \citep{Dea18}. At $x = 0$, a reflecting boundary condition is imposed, representing a rigid wall. The plasma, initially moving toward the left boundary ($-x$ direction), reflects upon encountering the wall. The interaction between the incoming and reflected plasma generates a relativistic perpendicular shock that propagates in the $+x$ direction.

\section{Simulation Results}
\label{sec_result}

The results of the 2D PIC simulation of the relativistic shock in the plasma composed of electrons and positrons with a proton admixture are presented in this section. We analyze the spatial structure of the shock, the electromagnetic fields, and the particle energy spectra, providing a discussion of the underlying physical mechanisms and their implications for particle acceleration.

\subsection{Shock Structure}

\begin{figure}
        \centering
        \includegraphics[width=0.5\textwidth]{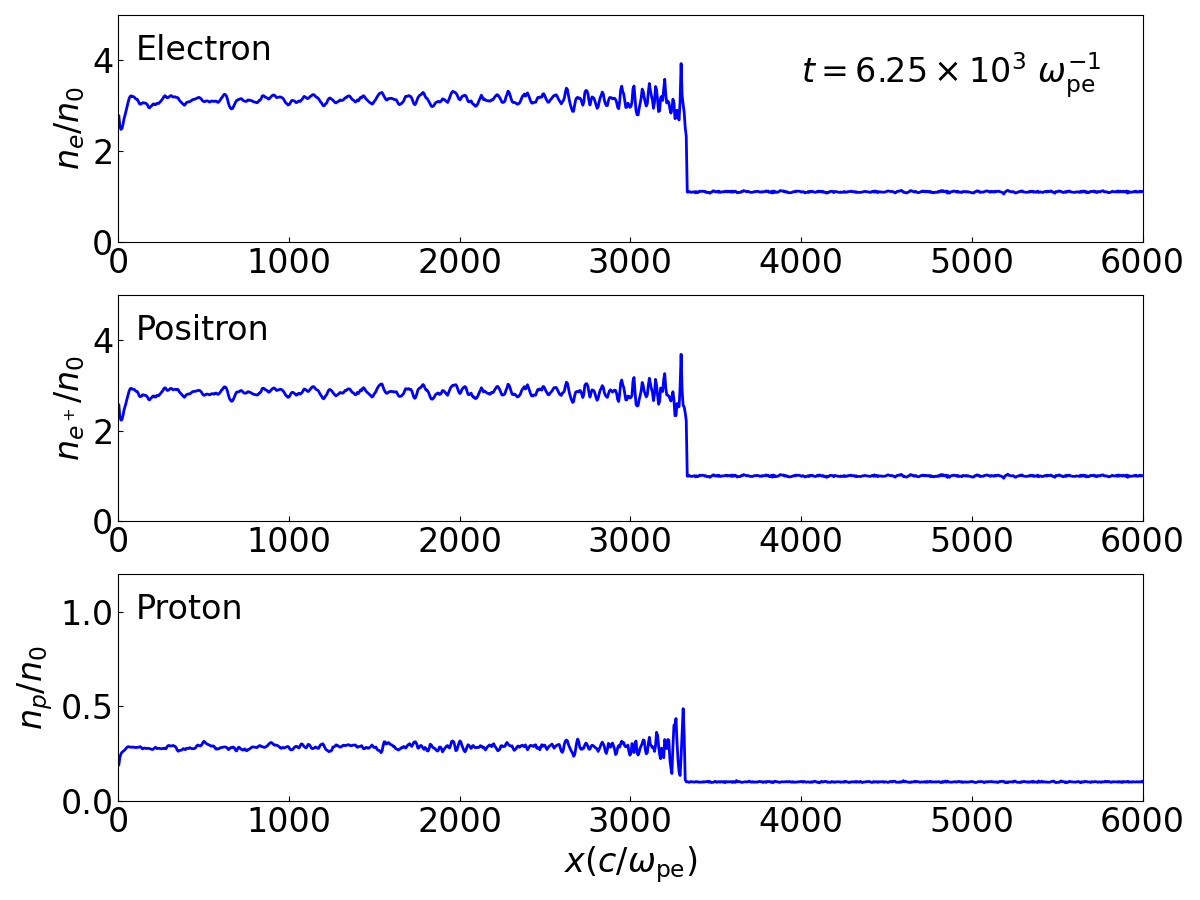}
        \caption{Particle number density normalized to $n_0$ as a function of $x$ at $t=6.25\times10^3 \omega_{\mathrm{pe}}^{-1}$.}
        \label{fig:density}
\end{figure}

\begin{figure}
        \centering
        \includegraphics[width=0.5\textwidth]{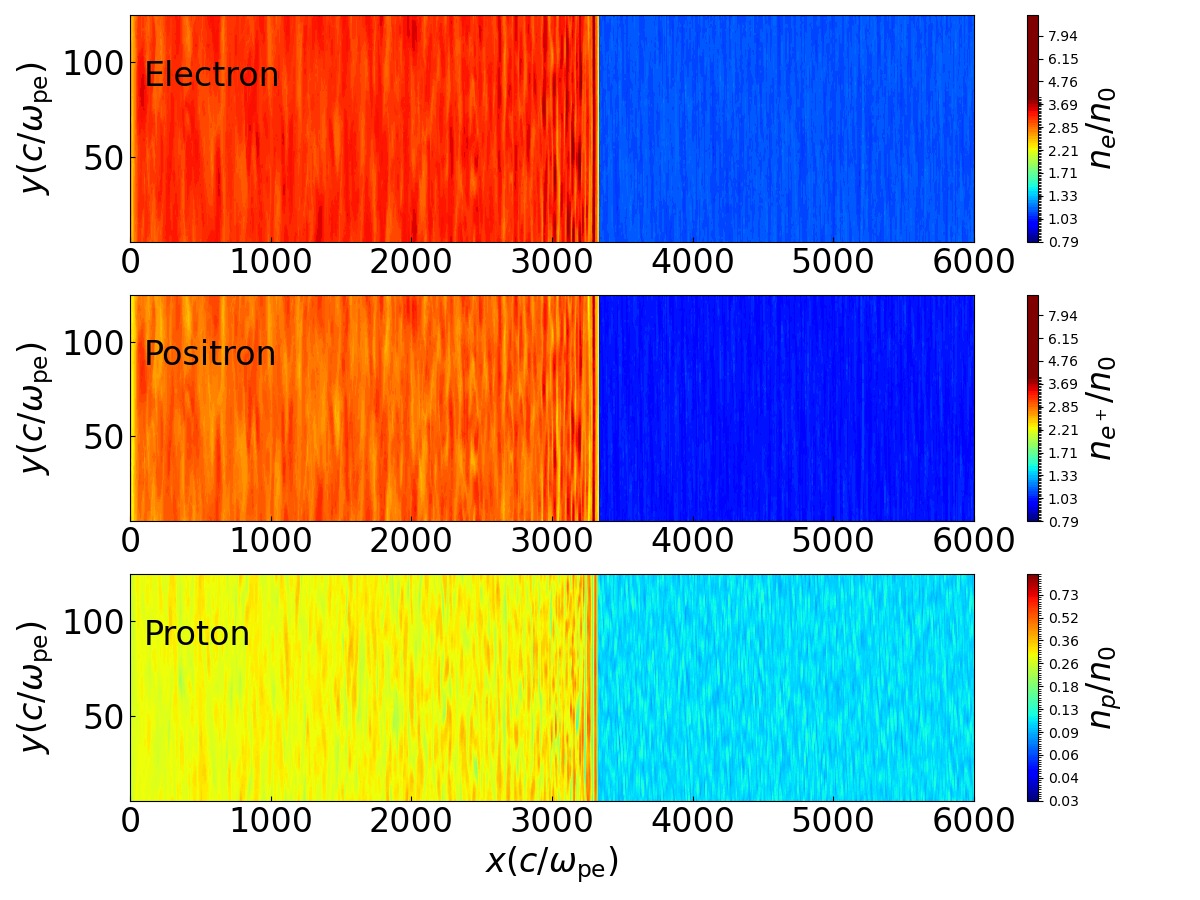}
        \caption{2D density distribution of electrons (top panel), positrons (middle panel), and protons (bottom panel) obtained from the simulation at $t=6.25\times10^3 \omega_{\mathrm{pe}}^{-1}$. The color scale indicates the relative density.}
        \label{fig:density_distribution}
\end{figure}

The spatial distribution of the electron, positron and proton number densities normalized to $n_0$ along the $x$-axis at  $t=6.25\times10^3 \omega_{\mathrm{pe}}^{-1} = 13.3\,\Omega_{\mathrm{Lp}}^{-1}$ are illustrated in Figure \ref{fig:density}. The 2D density distributions of electrons, positrons, and protons at $t=6.25\times10^3 \omega_{\mathrm{pe}}^{-1}$ are shown in Figure \ref{fig:density_distribution}. The shock front is located at $x = 3.3\times 10^3 \, c/\omega_{\rm pe}$, where the density exhibits a sharp increase, characteristic of a relativistic perpendicular shock. For electrons, positrons, and protons, the density ratio of particles far downstream to far upstream of the shock is $\sim 3$, consistent with theoretical predictions for 2D shocks. According to the conservation of mass, the shock position can be estimated to be located at $x_{\mathrm{sh}}= ct/(r_{\mathrm{ave}}-1)=3.48\times 10^3 \, c/\omega_{\rm pe}$ at $t=6.25\times10^3 \omega_{\mathrm{pe}}^{-1}$ with an averaged shock compression ratio of $r_{\mathrm{ave}}=2.9$. Downstream of the shock, the density profile shows significant fluctuations, which are attributed to the generation of turbulent magnetic fields. These fluctuations are a characteristic of relativistic shocks and play a crucial role in the acceleration and heating of particles.
The electron and positron densities are highly correlated, indicating strong coupling between the two species. This coupling is a result of the same mass-to-charge ratios, which cause them to respond similarly to the electromagnetic fields generated by the shock. The differential behavior of protons compared to electrons and positrons highlights the importance of considering the full plasma composition when studying relativistic shocks.

\subsection{Electromagnetic Field Structure}

\begin{figure}
        \centering
        \includegraphics[width=0.5\textwidth]{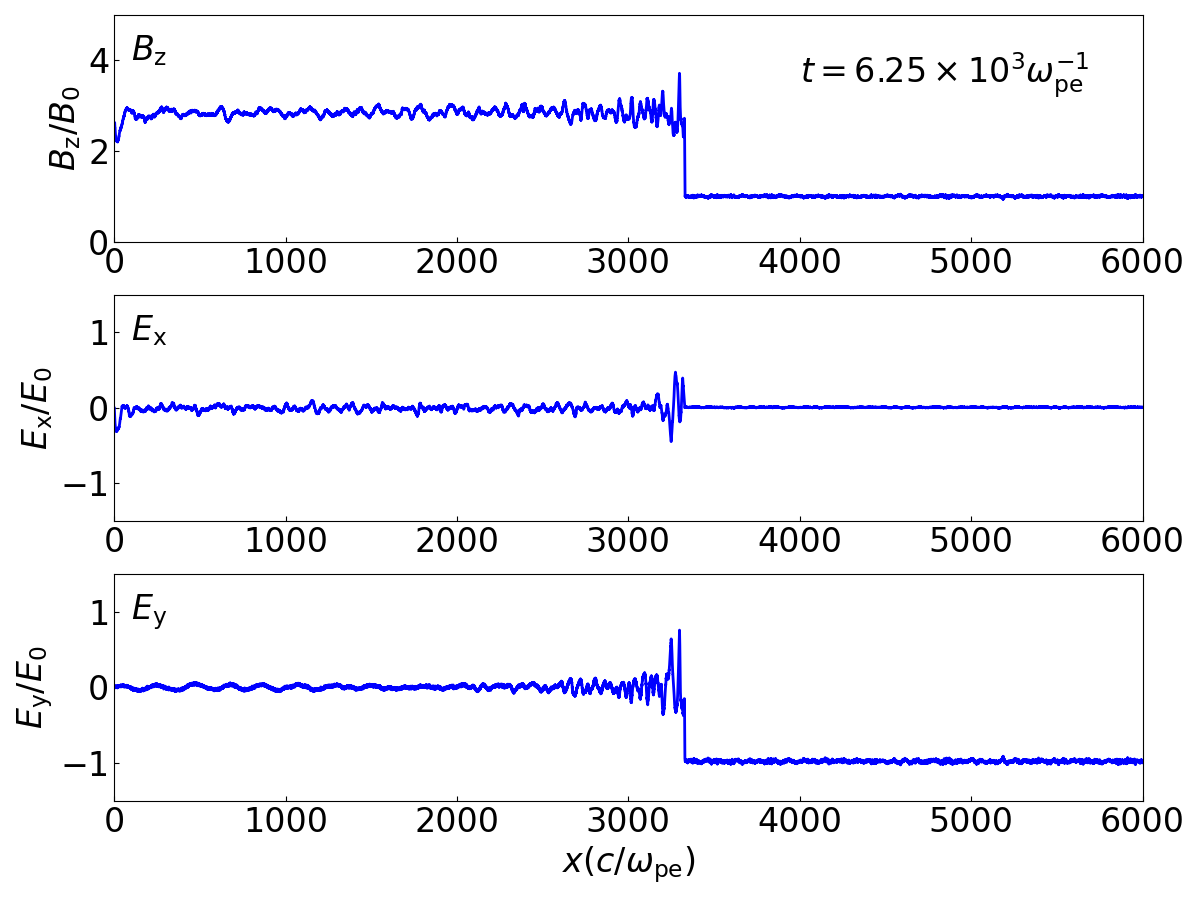}
        \caption{Spatial distribution of $B_z$, $E_x$ and $E_y$ at $t=6.25\times10^3 \omega_{\mathrm{pe}}^{-1}$. The magnetic field is amplified at the shock front and decays downstream.}
        \label{fig:bz}
\end{figure}

The magnetic field and electric field structures at $t=6.25\times10^3 \omega_{\mathrm{pe}}^{-1}$ are illustrated in Figure \ref{fig:bz}, which plots $B_z$, $E_x$ and $E_y$ as a function of $x$. The magnetic field exhibits fluctuations downstream of the shock, indicative of the turbulent environment generated by the shock. These fluctuations are driven by the Weibel instability, which is known to play a key role in the amplification of magnetic fields in relativistic shocks. The turbulent magnetic fields are essential for particle acceleration, as they provide the scattering centers necessary for the particle acceleration process. $E_x$ exhibits oscillations around zero, as expected for a transverse shock in a charge-neutral plasma.
$E_y$ transitions from its far-upstream value of $-E_0$ to approximately zero, with significant fluctuations in the downstream region near the shock.
These fluctuations of the electric field are driven by the motion of charged particles in the turbulent magnetic fields and play a key role in the acceleration and heating of particles in the shock.

\subsection{Particle Energy Spectra}

\begin{figure}
        \centering
        \includegraphics[width=0.5\textwidth]{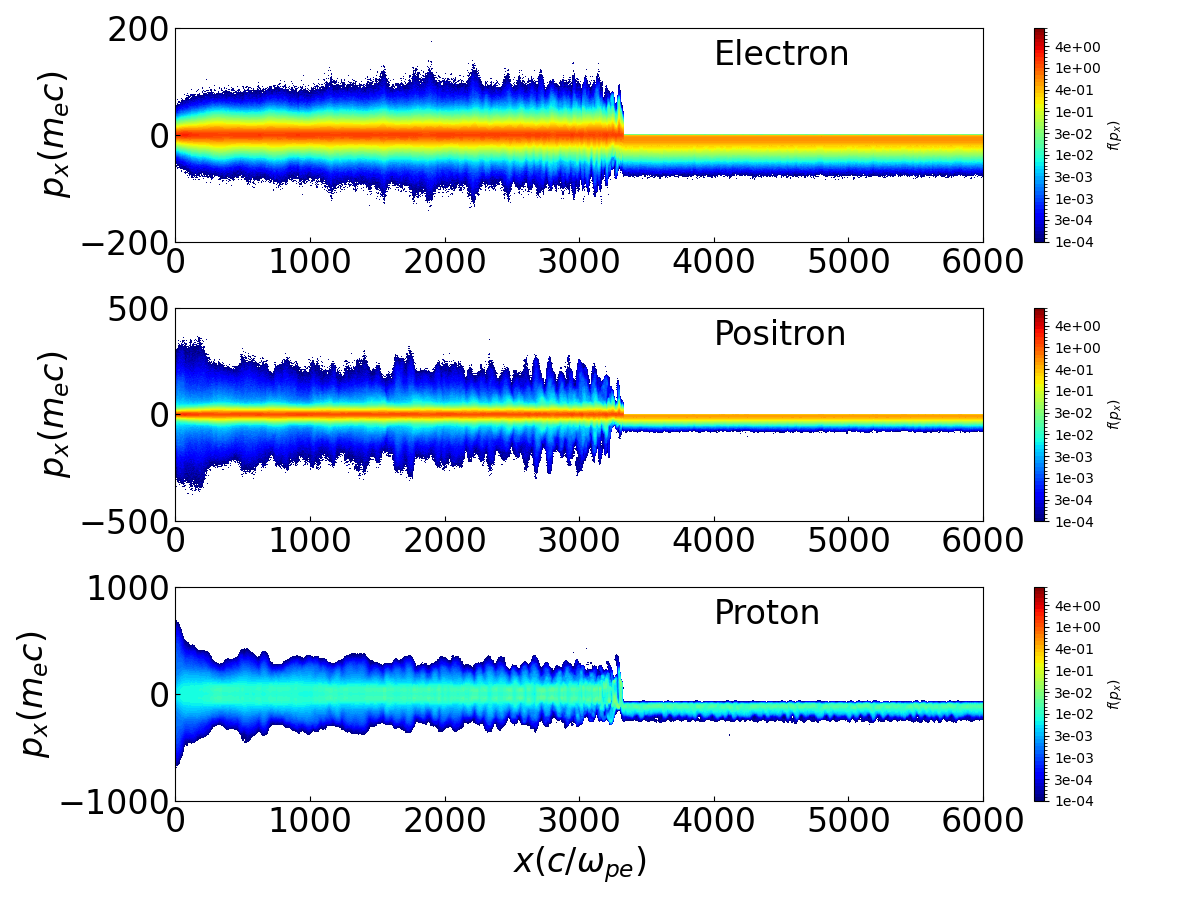}
        \caption{Longitudinal phase space plots $x - p_x$ of electrons, positrons, and protons at $t=6.25\times10^3 \omega_{\mathrm{pe}}^{-1}$. }
        \label{fig:phase}
\end{figure}

\begin{figure}
        \centering
        \includegraphics[width=0.5\textwidth]{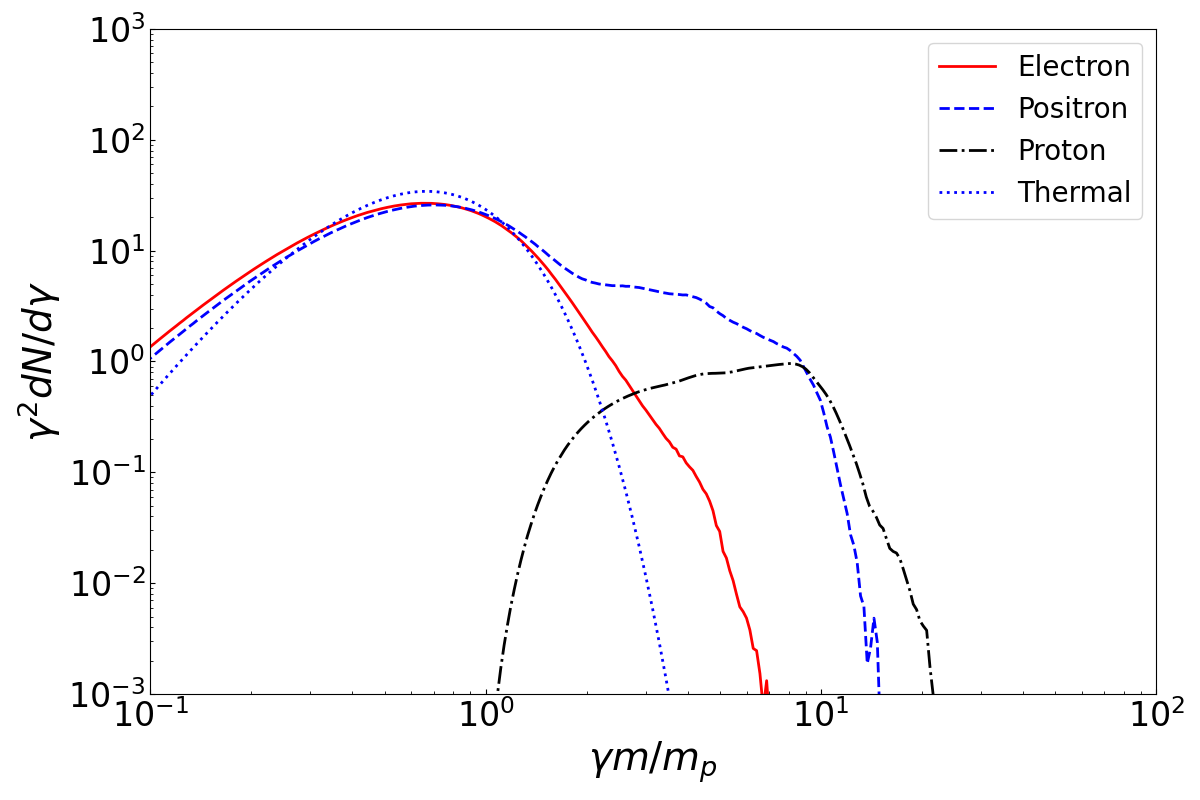}
        \caption{Energy spectra of the electrons, positrons, and protons downstream of the shock ($800\ c/\omega_{\mathrm{pe}}\leq x \leq 1600 \  c/\omega_{\mathrm{pe}}$) at $t=6.25\times10^3 \omega_{\mathrm{pe}}^{-1}$. The spectra consist of a thermal component at low energies and a nonthermal tail at high energies. The thermal distribution with a temperature of $T =5 m_e c^2/k_{\mathrm{B}}$ (dotted line) is also illustrated.}
        \label{fig:parspec}
\end{figure}

\begin{figure}
        \centering
        \includegraphics[width=0.5\textwidth]{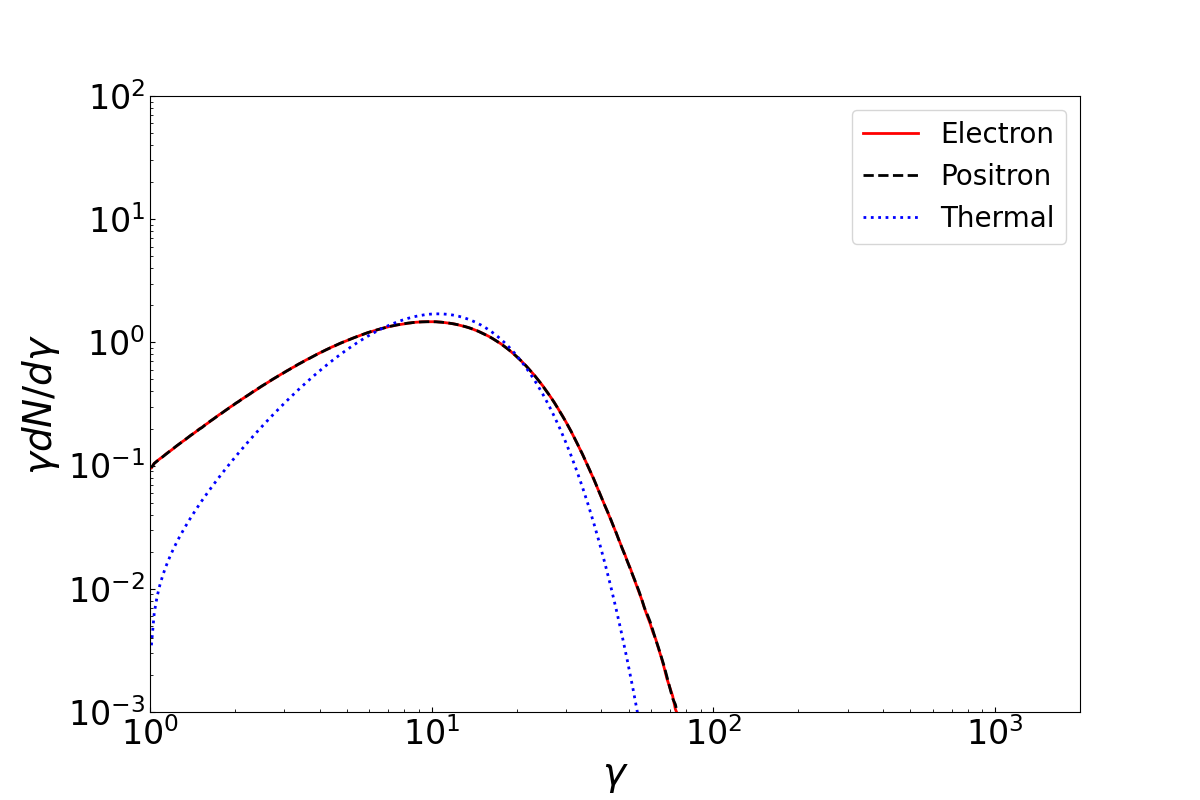}
        \caption{Energy spectra of the electrons and positrons downstream of the shock ($400\ c/\omega_{\mathrm{pe}}\leq x \leq 800 \  c/\omega_{\mathrm{pe}}$) at $t=1.75\times10^3 \omega_{\mathrm{pe}}^{-1}$ in a simulation without protons. The thermal distribution with a temperature of $T =3.5\, m_e c^2/k_{\mathrm{B}}$ (dotted line) is also illustrated.}
        \label{fig:parspecpe}
\end{figure}

\begin{figure}
        \centering
        \includegraphics[width=0.5\textwidth]{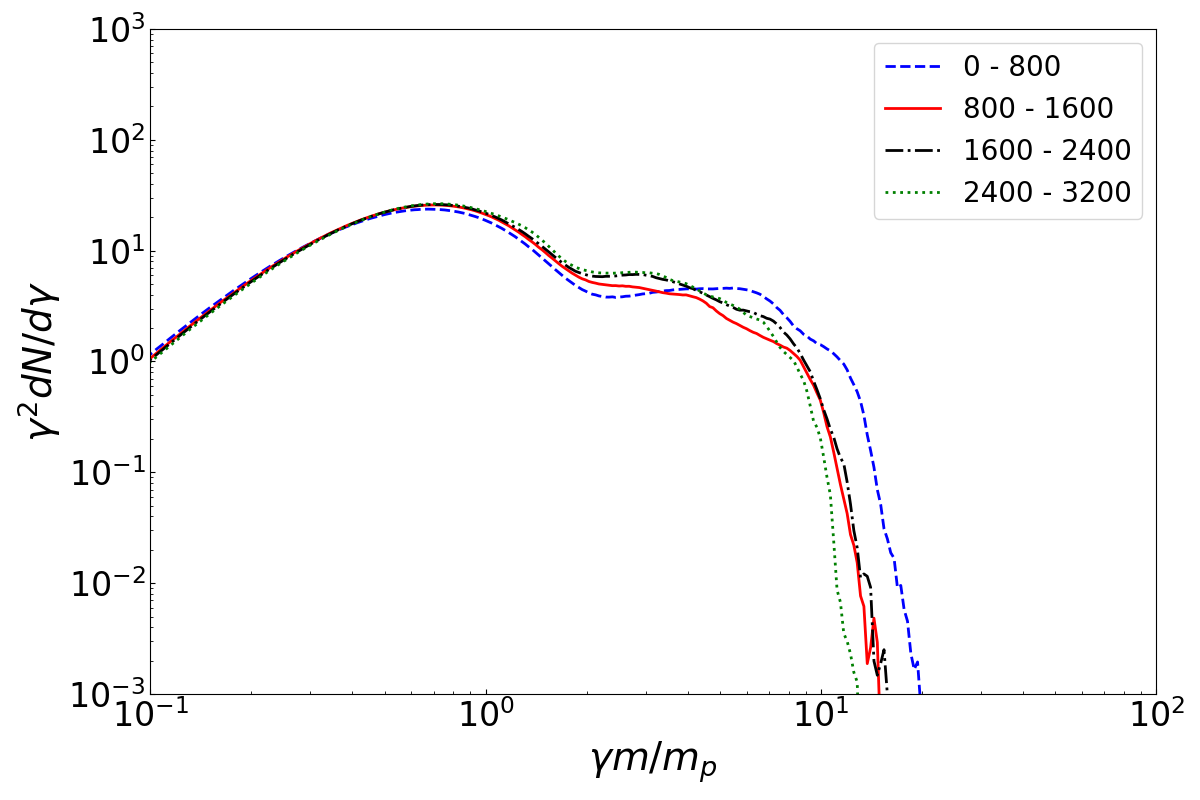}
        \caption{Energy spectra of the electrons, positrons, and protons in the downstream regions of the shock, i.e., $0\ c/\omega_{\mathrm{pe}} - 800 \  c/\omega_{\mathrm{pe}}$ (dashed line), $800\ c/\omega_{\mathrm{pe}} - 1600 \  c/\omega_{\mathrm{pe}}$ (solid line), $1600\ c/\omega_{\mathrm{pe}} - 2400 \  c/\omega_{\mathrm{pe}}$ (dash-dotted line), and $2400\ c/\omega_{\mathrm{pe}} - 3200 \  c/\omega_{\mathrm{pe}}$ (dotted line) at $t=6.25\times10^3 \omega_{\mathrm{pe}}^{-1}$.}
        \label{fig:PosiSpec_fenqu}
\end{figure}

\begin{figure}
        \centering
        \includegraphics[width=0.5\textwidth]{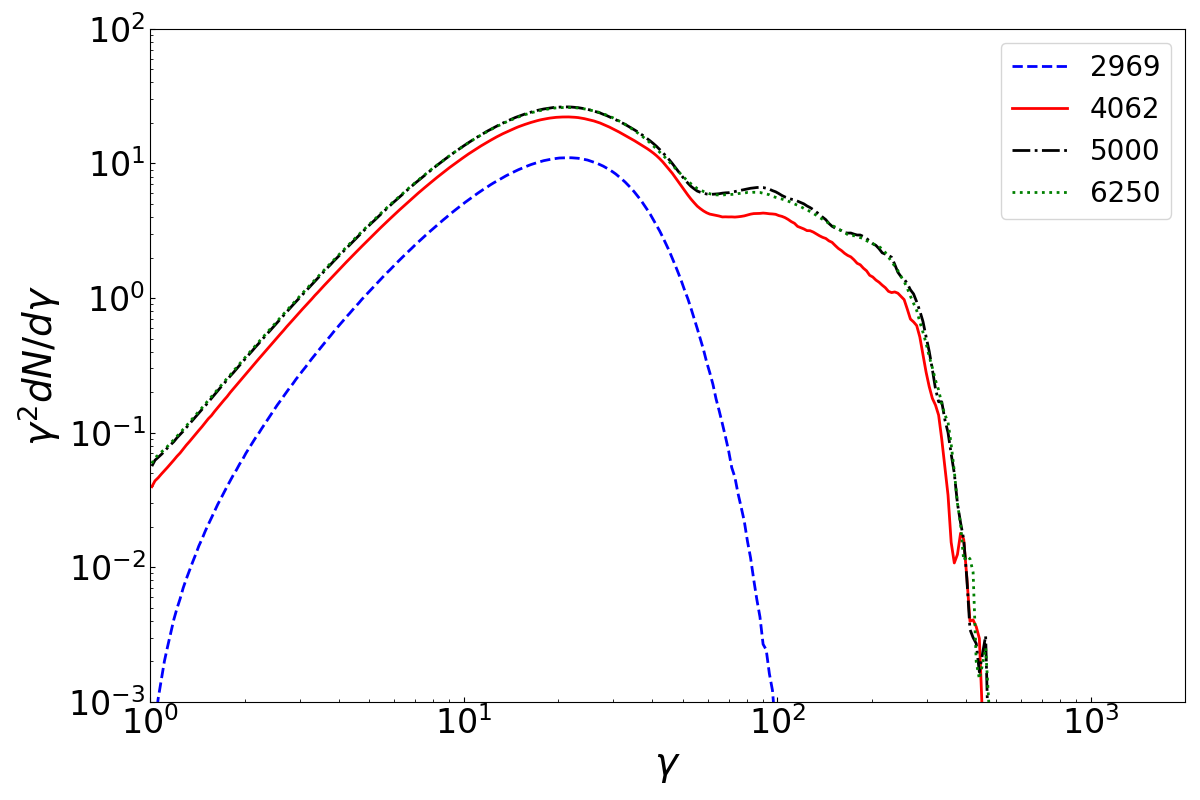}
        \caption{Energy spectra of the positrons in the downstream region of the shock ($200\ c/\omega_{\mathrm{pe}} - 800 \  c/\omega_{\mathrm{pe}}$) at $t=2969\, \omega_{\mathrm{pe}}^{-1}$ (dashed line), $4062\, \omega_{\mathrm{pe}}^{-1}$ (solid line), $5000\, \omega_{\mathrm{pe}}^{-1}$ (dash-dotted line), and $6250\, \omega_{\mathrm{pe}}^{-1}$, respectively.}
        \label{fig:Positime}
\end{figure}

Figure \ref{fig:phase} shows the numerically simulated phase-space distributions ($x - p_x$) of electrons, positrons, and protons. The incoming plasma is a cold plasma with a Lorentz factor of $10$, which gets heated after crossing the shock front. Compared to electrons, positrons exhibit a broader distribution in the phase space, indicating that they acquire greater momentum in the x-direction. The energy spectra of the electrons, positrons and protons downstream of the shock ($800\ c/\omega_{\mathrm{pe}}\leq x \leq 1600 \ c/\omega_{\mathrm{pe}}$) at $t=6.25\times10^3 \omega_{\mathrm{pe}}^{-1}$ are shown in Figure \ref{fig:parspec}. The energy spectrum of electrons slightly deviates from the thermal spectrum, while the protons have just been thermalized. The positrons downstream of the shock exhibit a thermal component at low energies, followed by a prominent nonthermal tail, and the maximum energy obtained by positrons is comparable to that of the protons. For comparison, we indicated particle acceleration in relativistic shocks in electron-positron plasmas without protons with the parameters same as the previous simulation with protons. Figure \ref{fig:parspecpe} shows the energy spectra of electrons and positrons downstream of the shock. A clear comparison with the proton-containing case reveals that neither electrons nor positrons undergo efficient acceleration in this scenario.

A striking feature of the particle spectra is the significant difference in the acceleration efficiency between positrons and electrons. As shown in Figure \ref{fig:parspec}, positrons are accelerated to energies comparable to those of protons, reaching Lorentz factors of $\gamma$ larger than $300$. In contrast, electrons are not efficiently accelerated and remain confined to lower energies, with a much steeper spectrum at high energies. This asymmetry in the acceleration efficiency is due to the polarization properties of the electromagnetic waves generated by proton-driven instabilities. Positrons, being positively charged, resonate more effectively with the waves, leading to enhanced energy gain, while electrons experience weaker interactions with the turbulent fields.

Figure \ref{fig:PosiSpec_fenqu} shows the energy spectra of positrons in different downstream regions of the shock. In downstream regions farther from the shock, the positrons exhibit higher energies and harder spectra, which shows that positrons can undergo acceleration for a certain period after crossing the shock front, as they interact with waves downstream of the shock.



Figure \ref{fig:Positime} shows the time evolution of positron energy spectra in the downstream region ($200\ c/\omega_{\mathrm{pe}} < x < 800 \  c/\omega_{\mathrm{pe}}$). The results reveal that a hard nonthermal spectrum quickly stabilizes in the downstream region  with minimal subsequent positron acceleration or increase in maximum energy at $t=4062\, \omega_{\mathrm{pe}}^{-1}$, $5000\, \omega_{\mathrm{pe}}^{-1}$, and $6250\, \omega_{\mathrm{pe}}^{-1}$, respectively. This saturation is predicted by the quasi-linear theory of \citet{HA91} and confirmed in the shock simulations of \citet{Hea92}, where positron acceleration ceases once the proton-emitted waves are depleted through resonant absorption.

The acceleration process occurs through two-stage instabilities in magnetized shocks in plasmas composed of electrons, positrons, and protons. Assuming the particles in the plasma are gyrating around the uniform magnetic field lines, the initial stage involves rapid thermalization of $e^\pm$ via cyclotron emission at harmonics of $\Omega_{ce^\pm} = eB/m_e c\gamma$ ($\gamma$  is the particle Lorentz factor), producing relativistic Maxwellian distributions \citep{HA91}. Protons subsequently emit extraordinary (X) modes at harmonics $\omega \approx n\Omega_{\mathrm{Lp}}=eB/m_p c\gamma$, $n$ is the harmonic number). When $n \gtrsim m_p/m_e$, these waves become resonant with the positrons under the condition $\omega= \omega_{ce^+}/\gamma_{e^+} \approx n\Omega_{\mathrm{Lp}}$ ($\omega$ is the wave frequency, and $\omega_{ce^+}$ is the nonrelativistic positron cyclotron frequency), and the positrons, which gyrate in the same sense as the emitted waves produced by the protons, can absorb these high-harmonic modes  \citep{HA91,AA06}. When the energy of positrons increases, their cyclotron frequency decreases, allowing them to further absorb lower-frequency proton synchrotron radiation. Alternatively, Since electrons and protons have opposite gyromotion, the resonance condition for the electrons, i.e., $\omega = - |\omega_{ce^-}|/\gamma_{e^-}$, cannot be satisfied. As a result, the positrons can be more energized than the electrons.

Particle-in-cell simulations \citep{HA91,AA06} demonstrate that this mechanism can produce nonthermal spectra for the positrons with power-law form in energy space, i.e.,
\begin{equation}
    N(E)dE \propto E^{-\alpha}dE.
        \label{eq:spectrum}
\end{equation}
The power-law index $\alpha \approx 2$ emerges from quasi-linear equilibrium between wave emission and absorption. Gyrating protons generate broadband magnetosonic waves via the synchrotron maser instability, while positrons resonantly absorb these waves. Energy balance  produces the power-law spectrum for the positrons with an index of $\sim 2$ when including phase-space effects \citep{Hea92}.  The power-law index $\alpha$ of the nonthermal positron spectrum depends on the plasma composition and the magnetization parameter. Proton-dominated systems can produce harder spectra \citep{HA91,Hea92}. When the magnetic energy density is significantly lower than the particle kinetic energy, protons emit synchrotron waves with broader spectral distributions. Under these conditions, positrons can resonantly absorb more high-frequency waves, leading to the formation of harder power-law spectra\citep{Hea92}.

The maximum positron energy is constrained by the highest-frequency waves emitted by the protons. Since the initial energy of the protons is $\gamma_0 m_p c^2$, the emitted waves can reach frequencies up to high harmonics of the proton cyclotron frequency. When the positron's cyclotron frequency decreases to match the lowest-frequency waves ($\omega \sim \Omega_{cp}$), the positron reaches the maximum energy \citep{Hea92}
\begin{equation}
    \gamma_{\text{max}} \approx \gamma_p \frac{m_p}{m_{e}}.
\end{equation}


\begin{figure}
        \centering
        \includegraphics[width=0.5\textwidth]{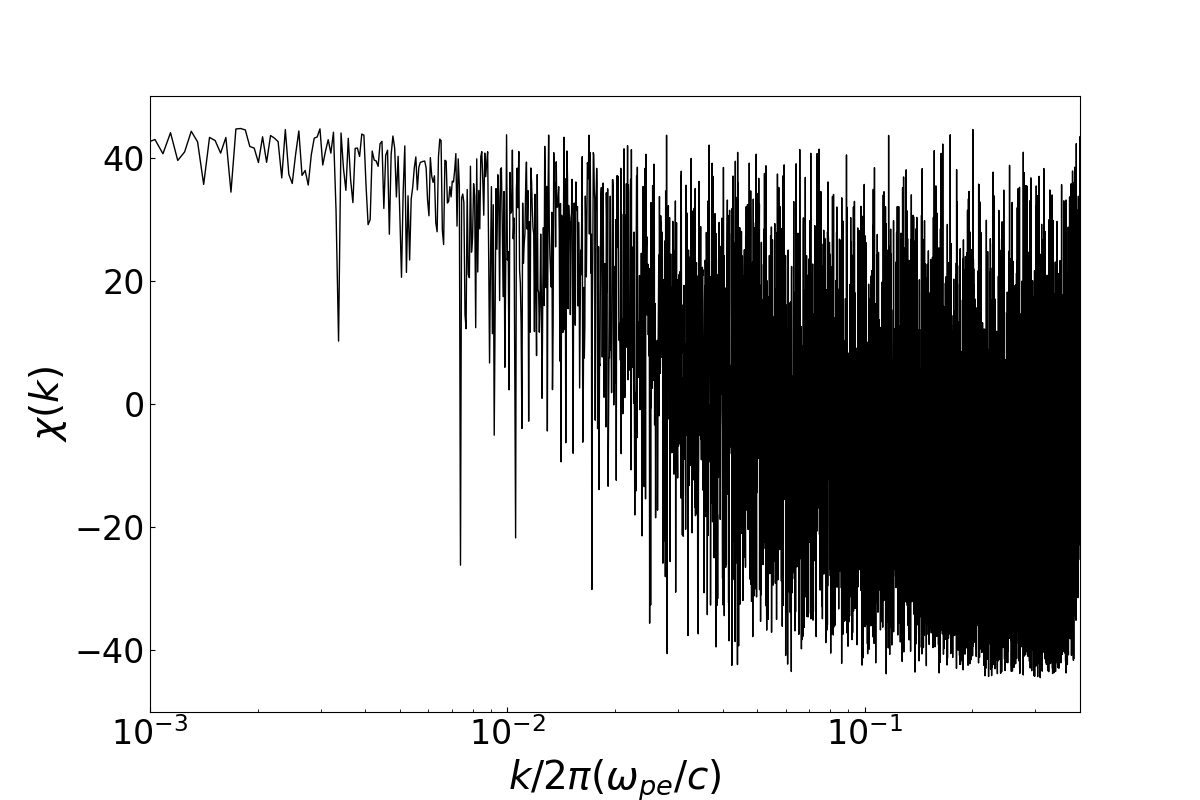}
        \caption{Spectral distribution of the polarization angle $\chi(k)$ in $k$ space for the magnetic field in the downstream ($1000\ c/\omega_{\mathrm{pe}}\leq x \leq 3300 \ c/\omega_{\mathrm{pe}}$) of the shock at $t=6.25\times10^3 \omega_{\mathrm{pe}}^{-1}$. }
        \label{fig:kai}
\end{figure}

The presence of protons, even at a low fraction ($\sim 5\%$ by number), significantly enhances the efficiency of nonthermal acceleration, particularly for positrons, which receive a larger share of the nonthermal energy compared to electrons. This asymmetry in the acceleration efficiency is consistent with the findings of \citet{AA06} and highlights the importance of ion-driven instabilities in shaping the particle spectra in relativistic shocks. The gyrating ions emit magnetosonic waves with left-handed orientation, and the waves facilitate the energization of the positrons compared with the electrons. The handedness of waves can be quantified by the polarization angle $\chi(k)=1/2 \sin^{-1}((|B_x(k)|^2 - |B_y(k)|^2)/(|B_x(k)|^2 + |B_y(k)|^2))$ in $k$ (wave number) space for the waves generated by the particles spinning around the magnetic field, and the positive (negative) value represent right-handed (left-handed) polarization. The waves generated by the protons with a Lorentz factor of $\gamma$ with frequency $\omega = m_p/m_e \Omega_{cp} = 0.2 (\gamma/10)^{-1} \omega_{pe}$ resonate with the positrons, which causes the positrons to be accelerated. As illustrated in Figure \ref{fig:kai}, the polarization angles near this frequency  are  negative, the waves are left-hand polarized, consistent with theoretical analysis.

\begin{figure}
        \centering
        \includegraphics[width=0.5\textwidth]{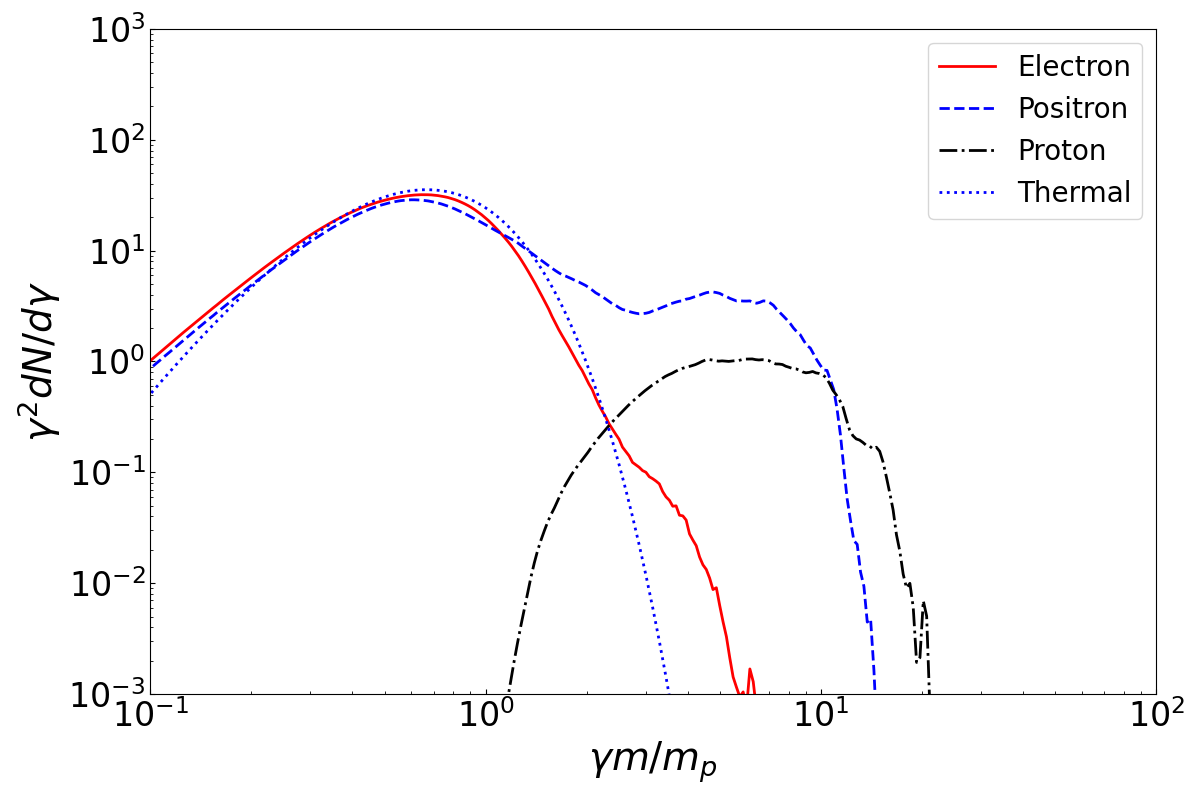}
        \caption{Energy spectra of the electrons, positrons, and protons downstream of the shock ($400\ c/\omega_{\mathrm{pe}}\leq x \leq 800 \  c/\omega_{\mathrm{pe}}$) at $t=1.72\times10^3 \omega_{\mathrm{pe}}^{-1}$ in the simulation with 64 particles per cell per species on a smaller spatial scale ($4096 \, c/\omega_{\mathrm{pe}}  \times 64 \, c/\omega_{\mathrm{pe}} $). The thermal distribution with a temperature of $T =5 m_e c^2/k_{\mathrm{B}}$ (dotted line) is also illustrated.}
        \label{fig:parspecpe64}
\end{figure}

To conserve computational resources, we employed $16$ particles per cell per species with a domain size of $12288 \, c/\omega_{\mathrm{pe}}  \times 128 \, c/\omega_{\mathrm{pe}}$ in the numerical simulations. To verify the reliability of the results, we performed additional calculations with 64 particles per cell per species on a smaller spatial scale ($4096 \, c/\omega_{\mathrm{pe}}  \times 64 \, c/\omega_{\mathrm{pe}} $). Figure \ref{fig:parspecpe64} shows the energy spectra of various particle species downstream of the shock ($400\ c/\omega_{\mathrm{pe}}\leq x \leq 800 \  c/\omega_{\mathrm{pe}}$) under these conditions at $t=1.72\times10^3\, \omega_{\mathrm{pe}}^{-1}$. It can be observed that positrons are accelerated more efficiently than electrons downstream of the shock, and the spectral characteristics are consistent with the former simulation using $16$ particles per cell per species.

The observed particle spectra and electromagnetic field structures provide valuable insights into the mechanisms of particle acceleration in relativistic shocks. The efficient acceleration of positrons, compared to electrons, suggests that the polarization properties of the electromagnetic waves play a crucial role in determining the energy gain of different particle species. The proton-driven instabilities generate waves that preferentially accelerate positrons, leading to the observed asymmetry in the energy spectra.

\section{Conclusions and Discussion}
\label{sec_cd}

We have performed 2D PIC simulations of the relativistic shock with a magnetization parameter of $\sigma=0.02$ in the plasma composed of electrons, positrons, and a small fraction of protons, focusing on the role of proton-driven instabilities in particle acceleration and the generation of nonthermal radiation. Our results provide new insights into the microphysics of relativistic shocks and their implications for astrophysical sources such as GRBs and PWNe.
The shock structure is characterized by fluctuating density and magnetic field. Downstream of the shock, the energy spectra of positrons consist of a thermal component and a nonthermal tail. A striking feature of the spectra is the significant difference in the acceleration efficiency between positrons and electrons: positrons are accelerated to energies comparable to those of protons, while electrons remain confined to lower energies. The positron energy spectrum in the far-downstream shock region displays a pronounced nonthermal component, reaching maximum energies similar to proton energies. This asymmetry is attributed to the polarization properties of the electromagnetic waves generated by proton-driven instabilities, which favor positron acceleration.

Earlier research on particle acceleration in relativistic shocks composed solely of electrons and positrons has demonstrated that efficient particle acceleration is observed in scenarios where the shocks are either weakly magnetized or parallel in configuration \citep{S08, SS09}. In weakly magnetized shocks, the reduced magnetic field strength allows for a more effective transfer of energy to the particles, facilitating their acceleration. Similarly, in parallel shocks, where the magnetic field lines are aligned with the shock normal, the absence of significant magnetic turbulence enables particles to be accelerated more efficiently along the field lines.

Our results are consistent with previous studies that have investigated particle acceleration in relativistic shocks. For example, \citet{Sea12} demonstrated that the presence of protons enhances the efficiency of nonthermal acceleration, particularly for positrons. Our work extends these findings by considering a plasma with a low fraction of protons, and the results suggest that the composition of the upstream plasma plays a critical role in determining the observed radiation properties. The presence of protons significantly enhances the efficiency of nonthermal acceleration, particularly for positrons. This asymmetry in the acceleration efficiency may help explain the observed properties of PWNe, where the presence of ions enables leptons to be accelerated within magnetized relativistic perpendicular shocks.

Although our simulations provide valuable insights into the microphysics of relativistic shocks, several important effects remain to be explored. For example, radiative cooling is not taken into account in the simulations. Furthermore, extending the simulations to three dimensions and longer timescales will help clarify the long-term evolution of the shock structure and particle acceleration processes. Although computational limitations prevented our simulation from running sufficiently long to achieve complete proton thermalization downstream, the positrons nevertheless develop a well-defined nonthermal spectrum, demonstrating robust acceleration. However, in the downstream region, positron energization shows no significant late-time enhancement, suggesting this process saturates when positrons reach energies comparable to the maximum energy of fully thermalized protons.

In conclusion, our results highlight the importance of ion-driven instabilities in shaping the particle spectra and radiation properties of relativistic shocks. These findings provide a foundation for future studies of astrophysical sources powered by relativistic shocks, and they underscore the need for self-consistent, multidimensional simulations to fully understand the complex interplay between particles and fields in these extreme environments.

\section*{acknowledgements}

This work was supported by the National Natural Science
Foundation of China (grant Nos. 12563010  and 12393852), Yunnan Fundamental Research Projects (grant No. 202501AS070068),  and Yunnan Revitalization Talent Support Program (XDYC-QNRC-2022-0486).

\bibliographystyle{aasjournal}
\bibliography{repep}{}
\bibliographystyle{aasjournal}
\end{document}